\documentclass[reprint,amsmath,amssymb,aps]{revtex4-1}
\usepackage{graphicx}
\usepackage{dcolumn}
\usepackage{bm}
\usepackage{hyperref}
\usepackage{color}

\begin{document}

\preprint{APS/123-QED}

\title{Thermal conductivity of bulk In$_{2}$O$_{3}$ single crystals}

\author{Liangcai Xu$^{1,2}$, Beno\^it Fauqu\'{e}$^{3}$, Zengwei Zhu$^{2}$, Zbigniew Galazka$^{4}$, Klaus Irmscher$^{4}$, Kamran Behnia$^{1}$}
 \affiliation{
 \makebox[\textwidth][c]{$^1$Laboratoire de Physique et Etude des Mat\'{e}riaux (UPMC-CNRS), ESPCI Paris, PSL Research University,75005 Paris, France}\\
 \makebox[\textwidth][c]{$^2$Wuhan National High Magnetic Field Center and School of Physics, Huazhong University of Science and Technology, Wuhan 430074, China}\\
 \makebox[\textwidth][c]{$^3$JEIP, USR 3573 CNRS, Coll\`ege de France, PSL Research University, 11, Place Marcelin Berthelot, 75231 Paris Cedex 05, France.}\\
 \makebox[\textwidth][c]{$^4$Leibniz-Institut f\"ur Kristallzüchtung, Max-Born-Str. 2, 12489 Berlin, Germany}\\
}

\date{\today}           

\begin{abstract}
The transparent semiconductor In$_{2}$O$_{3}$ is a technologically important material. It combines optical transparency in the visible frequency range and sizable electric conductivity. We present a study of thermal conductivity of In$_{2}$O$_{3}$ crystals  and find that around 20 K, it  peaks to a value as high as 5,000 WK$^{-1}$m$^{-1}$, comparable to the peak thermal conductivity in silicon and exceeded only by a handful of insulators. The amplitude of the peak drastically decreases in the presence of a type of disorder, which does not simply correlate with the density of mobile electrons. Annealing enhances the ceiling of the phonon mean free path. Samples with the highest thermal conductivity are those annealed in the presence of hydrogen. Above 100 K, thermal conductivity becomes sample independent.  In this intrinsic regime, dominated by phonon-phonon scattering, the magnitude of thermal diffusivity, $D$, becomes comparable to many other oxides, and its temperature dependence evolves towards  $T^{-1}$. The  ratio of $D$ to the square of sound velocity yields a scattering time which obeys the expected scaling with the Planckian time.

\end{abstract}

\maketitle
\section{INTRODUCTION}
In$_{2}$O$_{3}$ is a  technologically appealing wide-gap semiconductor (see~Ref.~\cite{Bierwagen_2015} for a review). Its main polymorph crystallizes in a cubic bixbyite structure~\cite{Karazhanov2007}. The cubic (primitive) unit cell of this structure contains 16 (8) molecules and 80 (40) atoms. Introducing Sn dopants turns this semiconducting solid to a transparent conductor known as indium tin oxide (ITO)~\cite{Walsh2010}. Electrical transport properties of doped In$_{2}$O$_{3}$ with the carrier density in the range 10$^{18}$ - 10$^{20}$ cm$^{-3}$  has documented a room-temperature mobility in the range carrier of $\sim$ 100-200 cm$^{2}$V$^{-1}$s$^{-1}$. The high transparency in the visible spectrum makes this system an established transparent conducting  material.  

Amorphous or granular indium oxide (InO$_x$) has been the subject of an independent fundamental research activity focusing on a fascinating two-dimensional superconductor~\cite{STEINER2005}. It is a platform to study metal-anomalous metal-superconductor transitions~\cite{Kapitulnik2019} and host to a number of remarkable low-temperature transport properties~\cite{Ovadia2009}. The connection between this two-dimensional electronic system and the superconducting ground state of crystalline doped thin films of In$_2$O$_3$~\cite{Makise2008} is yet to be clarified. Only recently~\cite{Hagleitner2012,Galazka2013a}, bulk single crystals of In$_2$O$_3$ became available. The thermal conductivitives of indium oxide thin films~\cite{Ashida2009} and single crystals ~\cite{Galazko2020} were previously reported at room temperature. Now, thanks to the direct growth of single crystals from the melt~\cite{Galazka2013a}, it is  possible to measure thermal conductivity of single crystals with a steady-state setup down to cryogenic temperatures. 

 \begin{table*}
 \centering
 \caption{ \textbf{Samples:} The In$_2$O$_3$ single crystals investigated by the present study. $n$ and $\rho$ refer to room temperature carrier concentration and resistivity. $\kappa_{peak}$ refers to the peak thermal conductivity. All samples were cut to $5\times 3\times 0.5$ mm$^3$. The orientations relate to the largest plane (5 $\times$ 3 mm$^2$). }
 \begin{ruledtabular}
\begin{tabular}{lccccc}
Sample no. & Orientation & Annealing conditions & $n$(10$^{17}$cm$^{-3}$)& $\rho$(m$\Omega$cm)  & $\kappa_{peak}$ (WK$^{-1}$m$^{-1}$)\\
\hline
1 & (111) & As grown & 12 & 17  & 100  \\
2 & (111)& 900$^{o}$C, 40 h, air & 2 & 95 & 300 \\
3 & (001) & 1000$^{o}$C, 40 h, air & 3 & 89  & 450 \\
4 & (110) & 1000$^{o}$C, 40 h, O$_2$ & 12  & 45 & 480  \\
5 & (111) & 1000$^{o}$C, 40 h, O$_2$ and 650$^{o}$C, 10 h, H$_2$ (5\%)+Ar & 180  & 2.1 & 1200  \\
6 & (111) & 1000$^{o}$C, 40 h, O$_2$ and 500$^{o}$C, 10 h, H$_2$ (5\%)+Ar & 61  & 10 &  5000  \\
\hline
\end{tabular}
\end{ruledtabular}
\label{Tab1}
\end{table*}
\begin{figure}
    \centering
    \includegraphics[width=6cm]{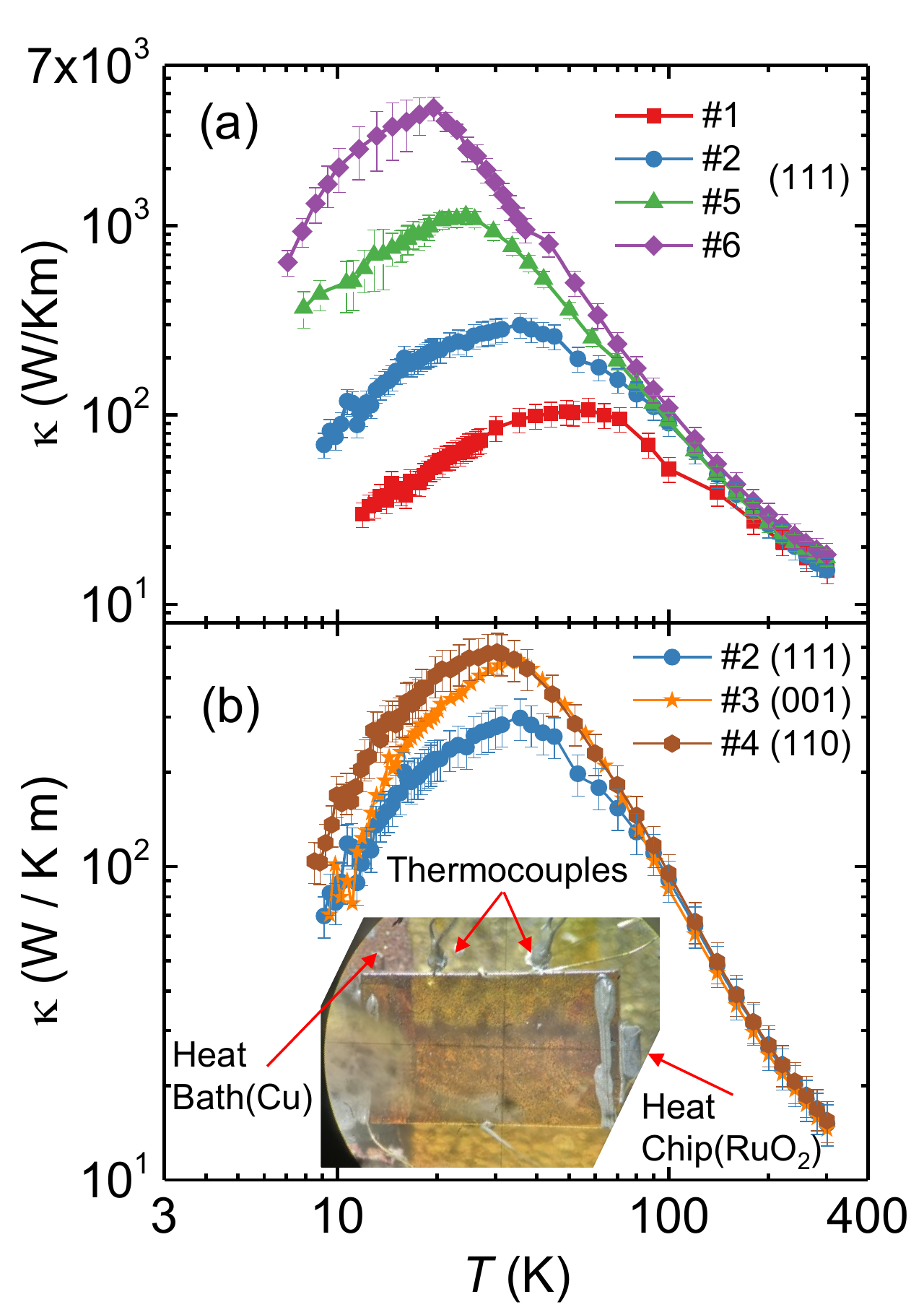}
    \caption{Thermal conductivity of In$_{2}$O$_{3}$ and its sample dependence. (a) Thermal conductivity $\kappa$ of four different samples of In$_{2}$O$_{3}$. Note the strong variation in the magnitude of maximum $\kappa$ among the samples.  Purple bars represent the margin of experimental reproducibility in repeated measurements on sample 6. Because of its large thermal conductance, the temperature difference caused by the application of the heat current was small, leading to a larger experimental uncertainty. (b) Comparison of thermal conductivity in three samples having different plane orientations.  The heat current was always applied along the longest axis. No anisotropy is expected in this cubic system and the difference is due to different levels of disorder. The inset shows the measurement setup with thermocouples.}
    \label{fig:Kappa_samples}
\end{figure}


In this paper, we present a study of temperature dependence of thermal conductivity in single crystals of In$_{2}$O$_{3}$.  We find that the peak thermal conductivity in In$_{2}$O$_{3}$ crystal can reach as high as 5000 WK$^{-1}$m$^{-1}$. In a few other solids, such as diamond~\cite{Onn1992,Inyushkin2018} and sapphire (crystalline Al$_2$O$_3$)~\cite{Cahill1988}, this peak is exceeded in magnitude. Most oxides show a much more modest maximum thermal conductivity. We find that this peak is drastically reduced by disorder. However, the level of disorder is not simply set by dopant concentration and depends on annealing conditions. Samples annealed in the presence of hydrogen are those showing the highest thermal conductivity. Above 100 K, thermal conductivity $\kappa$ in all samples converges to a similar magnitude. In this intrinsic regime (governed by phonon-phonon scattering), the ratio of the thermal diffusivity to the square of sound velocity respects a boundary observed recently for all insulating solids~\cite{Martelli2018,Behnia_2019}. Comparison with Al$_2$O$_3$ indicates that, in agreement with a picture recently proposed to explain this boundary,  the compound with the lighter atom has a larger thermal diffusivity driven by its superior sound velocity~\cite{Mousatov2020}. 

\section{EXPERIMENTAL}
Bulk In$_2$O$_3$ single crystals were grown with a crystal growth technique under the name levitation-assisted self-seeding crystal growth method (LASSCGM) as described in detail by Galazka \textit{et al.} \cite{Galazka2013a,Galazka2014,Galazko2020} that enabled to obtain bulk single crystals up to 35 mm in diameter and 12 mm thick. The as-grown crystals were dark red or dark brown and highly conducting with a free electron concentration of (1-4)$\times 10^{18}$ cm$^{-3}$ and Hall electron mobility of 130-170 cm$^2$V$^{-1}$s$^{-1}$~\cite{Galazka2014,Galazko2020}. Annealing in the presence of oxygen and hydrogen allowed tuning the free electron concentration between 1$\times 10^{18}$ and  6$\times 10^{19}$ cm$^{-3}$ and increased the room-temperature Hall electron mobility to  200 cm$^2$V$^{-1}$s$^{-1}$. Wafers fabricated from bulk single crystals were characterized by a very narrow rocking with the full width at half maximum (FWHM) $<$ 50 arcsec and the x-ray texture indicating no grain boundaries~\cite{Galazka2013a,Galazka2014}. The dislocation etch pit density was found to be between 5$\times 10^4$ and 3$\times 10^5$ cm$^{-2}$~\cite{Galazka2014,Galazko2020}. The  crystal orientation, the annealing conditions, and the size of the single crystals studied in this work are listed in Table \ref{Tab1}. All crystals had all identical dimensions.

We measured thermal conductivity in a steady-state configuration with one heater and two thermometers. Like in a setup used  before~\cite{Xu2020}, we used chromel-constantan thermocouples  as thermometers between 10 and 300 K [see inset in Fig. 1(b)]. In the case of sample 6, the data were completed by measurements using Cernox chips as thermometers between 2 and 30 K. The two sets of data were found to be consistent at intermediate temperatures. 

We can identify two potential sources of experimental uncertainty. The first is the finite size of the thermal contacts. In this case, the sample length was 5 mm and the  typical contact size was 0.1 mm, implying an uncertainty of 2\% on the absolute value. The other source of uncertainty is heat loss along the measuring wires. The geometric factor of our measuring wires, given the thickness (25 $\mu m$) and their length ($\simeq$15 mm), is three orders of magnitude smaller than the geometric factor of our samples. Thermally resistant manganin wires were used to connect thermocouples to the heat bath, and therefore, given manganin thermal conductivity ($\kappa_{manganin}\approx 1$ W/Km at 10K and $\approx 20$ W/Km at  300K~\cite{Touloukian1970}) the margin due to heat loss is less than ~3\%  of the sample thermal conductivity.

\begin{figure}[htb]
    \centering
    \includegraphics[width=8.0cm]{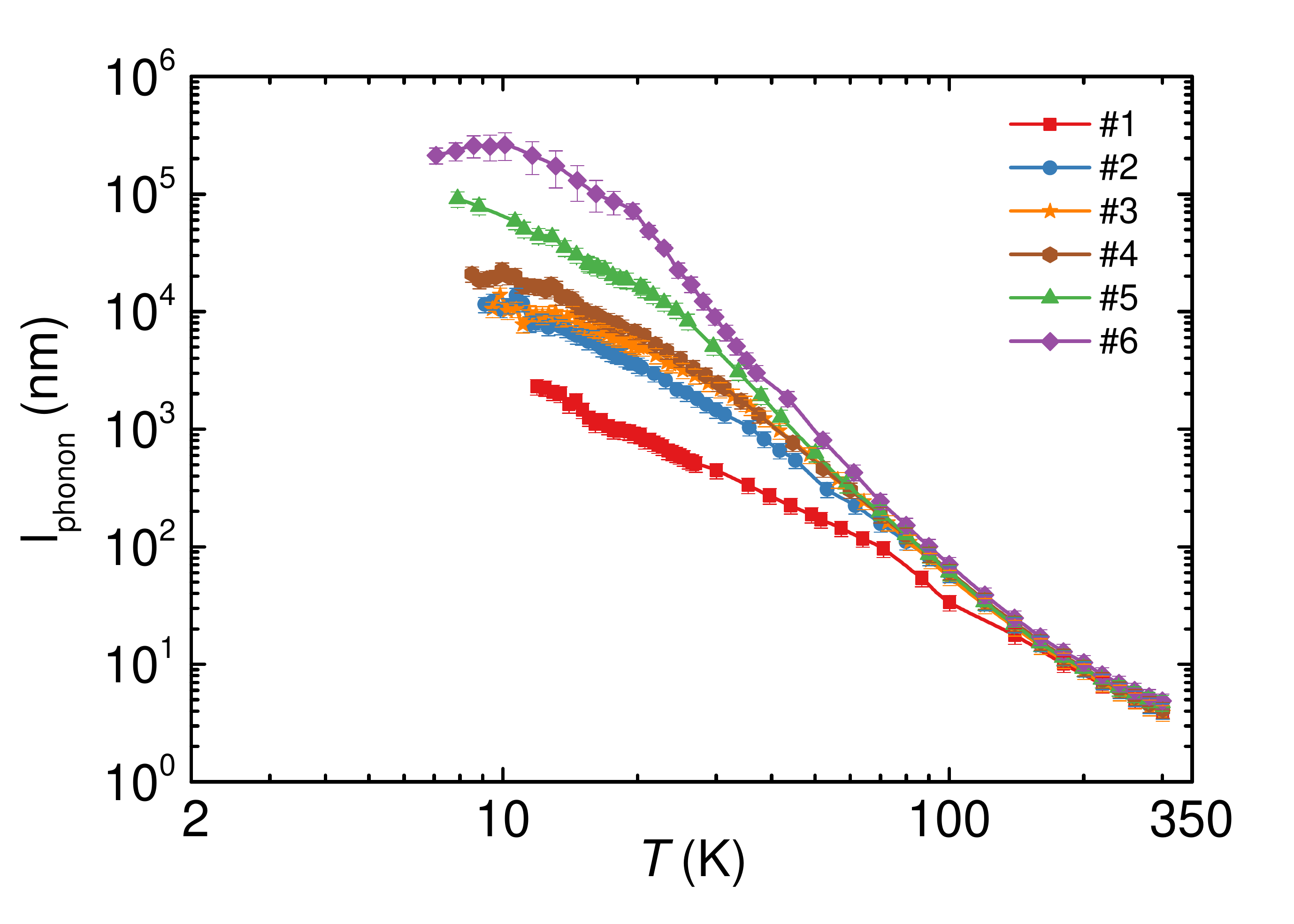}
    \caption{Phonon mean free path in In$_{2}$O$_{3}$, extracted from thermal conductivity, specific heat, and the sound velocity. The electric resistivity and the carrier density of these samples are listed in Table \ref{Tab1}. There is no visible correlation between the electronic properties and phonon mean free path. Note that in the best sample (6), the mean free path approaches the sample thickness of 0.5 mm at low temperature.}
    \label{fig:Basic_properities}
\end{figure}
\section{RESULTS}
Figure \ref{fig:Kappa_samples} presents our main results. Near room temperature, thermal conductivity increases with cooling and its magnitude is the same in all crystals studied. The room-temperature thermal conductivity of our bulk samples, $\kappa$ (300 K) = 15$\pm$2 WK$^{-1}$m$^{-1}$, is comparable to the value of 13.1 WK$^{-1}$m$^{-1}$ obtained by the laser flash method~\cite{Galazko2020} and three times larger than what was reported for thin films of ITO ($\approx$ 5 WK$^{-1}$m$^{-1}$)~\cite{Ashida2009}. In all samples,  $\kappa (T)$ shows a broad peak. The magnitude of this $\kappa_{peak}$ is strongly sample dependent. It varies from 100 WK$^{-1}$m$^{-1}$ in sample 1 to 5000 WK$^{-1}$m$^{-1}$ in sample 6. The higher the $\kappa_{peak}$, the lower the temperature at which it occurs. These features point to a smooth evolution in disorder among these samples. As we see below, however, identifying this disorder is not trivial.

We measured the temperature dependence of resistivity and  Hall constant $R_{H}$, of all samples. Carrier density was determined using $n = 1/R_{H}e$ ($e$ is the elementary charge). This single-band expression is justified by the fact that the minimum of the conduction band is at the center of the Brillouin zone~\cite{Bierwagen_2015}. The room-temperature values are given in Table \ref{Tab1}.  

Mobile electrons can also conduct heat. In general, the measured thermal conductivity has two contributions: $\kappa_{meas} = \kappa_{ph} + \kappa_{e}$. The latter is the electronic contribution to thermal conductivity, which can be estimated using the electric resistivity and the Wiedemann-Franz law: $\kappa_{e} = L_0T/\rho$, where $L_0=2.45 \times 10^{-8}$ V$^2$/$K^2$ is the Sommerfeld value. Even in the most metallic like sample, $\kappa_{e} \ll \kappa_{meas}$. Indeed, for $\rho \approx 2 m\Omega$cm, one expects $\kappa_{e}/T$=$L_0/\rho\approx 10^{-3}$ WK$^{-2}$m$^{-1}$. Even at room temperature, this is a tiny fraction of the measured signal.  We can therefore safely conclude that $\kappa_{meas} \approx \kappa_{ph}$.

The effective phonon mean free path $\ell_{ph}$ can be estimated using the kinetic equation $\kappa = \frac{1}{3}~C_{ph}v_{s}\ell_{ph}$. (Note that such an estimation neglects the possible variation of the mean free path among modes and frequencies.) Figure \ref{fig:Basic_properities} shows the temperature dependence of the effective phonon mean free path $\ell_{ph}$ extracted from our thermal conductivity data, the reported specific heat~\cite{Cordfunke1992,bachmann1981low}), and the estimated sound velocity $v_{s}$ = 4.4 km/s \cite{Ashida2009}.
The smooth evolution of $\ell_{ph} (T)$ (Fig. \ref{fig:Basic_properities}) is reminiscent of the way thermal conductivity of an insulator evolves upon controlled introduction of disorder, such as NaCl crystals with Ag colloids~\cite{Vandersande}. Surprisingly, however, the smooth evolution of $\ell_{ph} (T)$ is not correlated with the electrical properties. As an example, the sample showing the longest $\ell_{ph}$ (sample 6) and the one showing the shortest (the as-grown sample, sample 1) are only slightly different from the charge transport viewpoint. The  room-temperature carrier density in sample 6 is five times larger than in sample 1. How can then the phonon mean free path in sample 6 be a hundred times longer, in spite of having a higher concentration of scattering centers?
\section{DISCUSSION}
The behavior seen here is to be contrasted with the case of doped semiconductors such as silicon and germanium~\cite{carruthers1957} or doped oxides such as SrTi$_{1-x}$Nb$_{x}$O$_3$~\cite{Martelli2018}, where the introduction of Nb dopants leads to a steady enhancement in electrical conductivity (because of the introduction of mobile electrons) and a steady decrease in thermal conductivity (because of a rise in the number of the scattering centers). 

Our observation implies that the case of In$_{2}$O$_{3}$ is more complicated and disorder length scale is not simply the interdopant distance. In the as-grown sample sample 1, the deviation from the intrinsic regime starts below 150 K (see Fig. \ref{fig:Basic_properities}). The typical wavelength of an acoustic phonon at this temperature is $\lambda_{ph}$ (150 K)=$\frac{hv_s}{k_BT}$=1.4 nm. On the other hand, in sample 2, the deviation starts at 30 K, when the phonon wavelength is $\lambda_{ph}$(30 K) = 7 nm.  These numbers set the lower boundary to the size of scattering centers. The most prominent candidates for playing this role  are clusters of oxygen vacancies which manifest themselves as In particles. They are formed during melt growth in the cooling crystal and have been identified in a recent study combining electron microscopy, light scattering techniques, and optical absorption spectroscopy ~\cite{Albrecht}. The concentration of these clusters, which can be as large as several tens of nanometers~\cite{Albrecht}, diminishes upon annealing. This type of extended disorder is the most plausible candidate for scattering long wave-length acoustic phonons in low-$\kappa$ samples.

The two samples with highest $\kappa_{peak}$ were annealed in the presence of H$_2$ (see Table \ref{Tab1}). Hydrogen is a possible dopant and a shallow donor impurity~\cite{Limpi2009}.  H atoms can replace O atoms at their site~\cite{Galazka2013b}. The small size of H dopants may allow the survival of high phonon thermal conductivity (which requires small disorder) despite finite electric conductivity (which requires atomic substitution). Note, however, that according to first-principles calculations~\cite{Varley_2011}, hydrogen may also occupy interstitial positions and form complexes with indium vacancies. More extensive studies are required to pin down the origin of the link between annealing conditions and peak thermal conductivity. Note that in the most heat conducting , the phonon mean free path $\ell_{ph} (T)$ becomes as long as 0.3 mm below 10 K, only slightly shorter than the sample thickness (0.5 mm), implying that phonons are close to ballistic. 

A theoretical description of the thermal conductivity can be achieved by calculating the complex phonon spectrum of this solid  and solving the Boltzmann-Peierls transport equation~\cite{Lindsay2016}. In the absence of a rigorous solution, there are  approximate treatments. The most influential is the one introduced by Callaway~\cite{Callaway1959}, which in contrast to relaxation time approximation takes into account normal collisions between phonons~\cite{Allen2013}. The starting point for a phenomenological approach is the Debye temperature. However, because of the complex phonon spectrum due to the large number of atoms per unit cell, a single well-defined Debye temperature is absent in this system. Published Debye temperatures of In$_2$O$_3$ range from 420 to 811 K (see Ref. \cite{Preissler2013} for more details on this issue). In a forthcoming paper, Subedi \cite{Subedi} will present \textit{ab initio} calculations of the phonon spectrum and lattice thermal conductivity. His results give a reasonable account of our experimental data for the cleanest samples in a wide  temperature window.
\begin{figure}
    \centering
    \includegraphics[width=8.5cm]{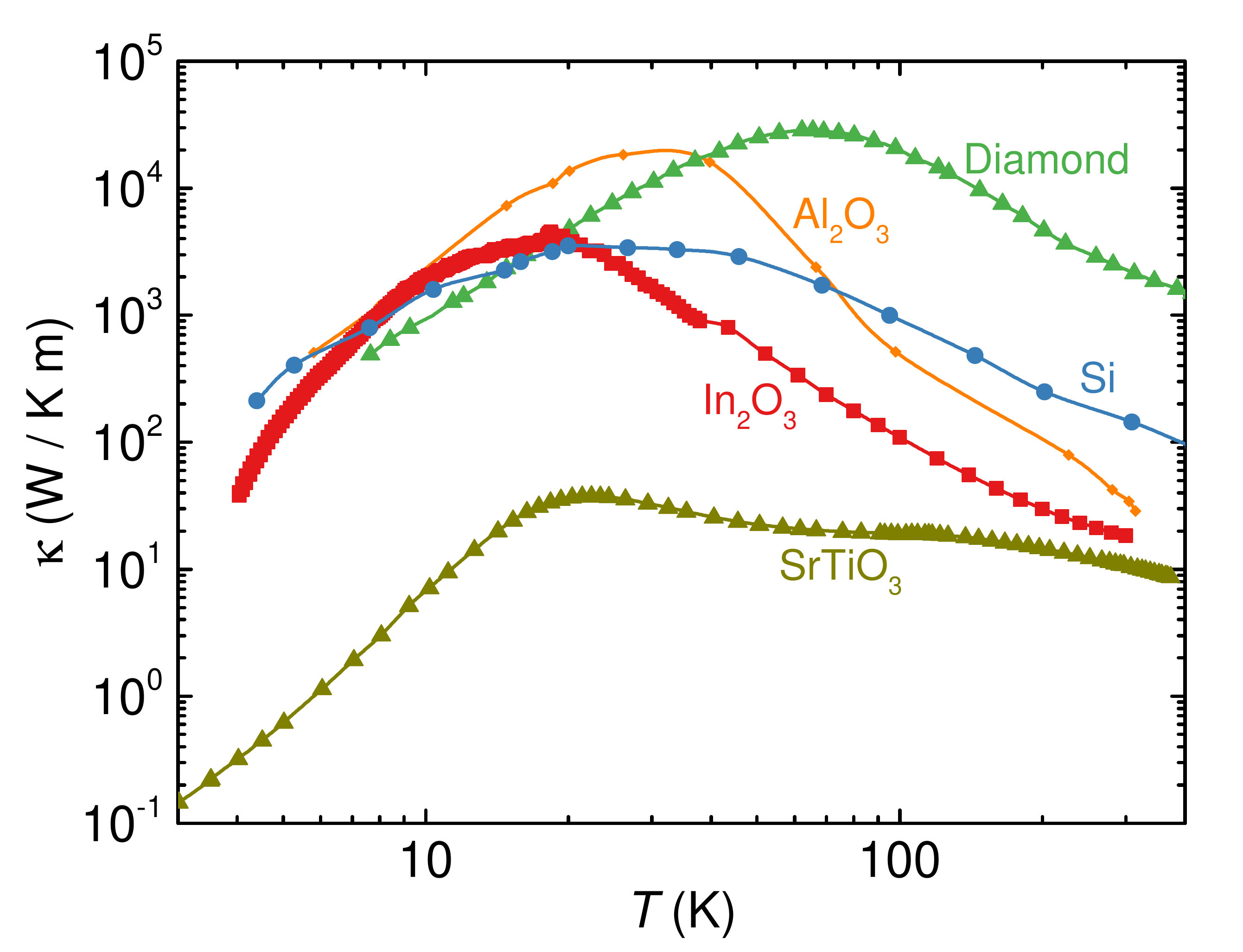}
    \caption{\textbf{Comparison with other solids}. Comparison of the thermal conductivity in the best In$_{2}$O$_{3}$ sample with  diamond~\cite{Inyushkin2018}, silicon~\cite{Glassbrenner1964}, Al$_{2}$O$_{3}$ (sapphire)~\cite{Cahill1988}, and a perovskite  SrTiO$_{3}$ \cite{Martelli2018}. Thermal conductivity in In$_{2}$O$_{3}$ is comparable to a ternary oxide such as SrTiO$_{3}$ at room temperature, but peaks to a value comparable to silicon's peak at low temperature.}
    \label{fig:Kappa_campare}
\end{figure}

\begin{figure}
    \includegraphics[width=8.5cm]{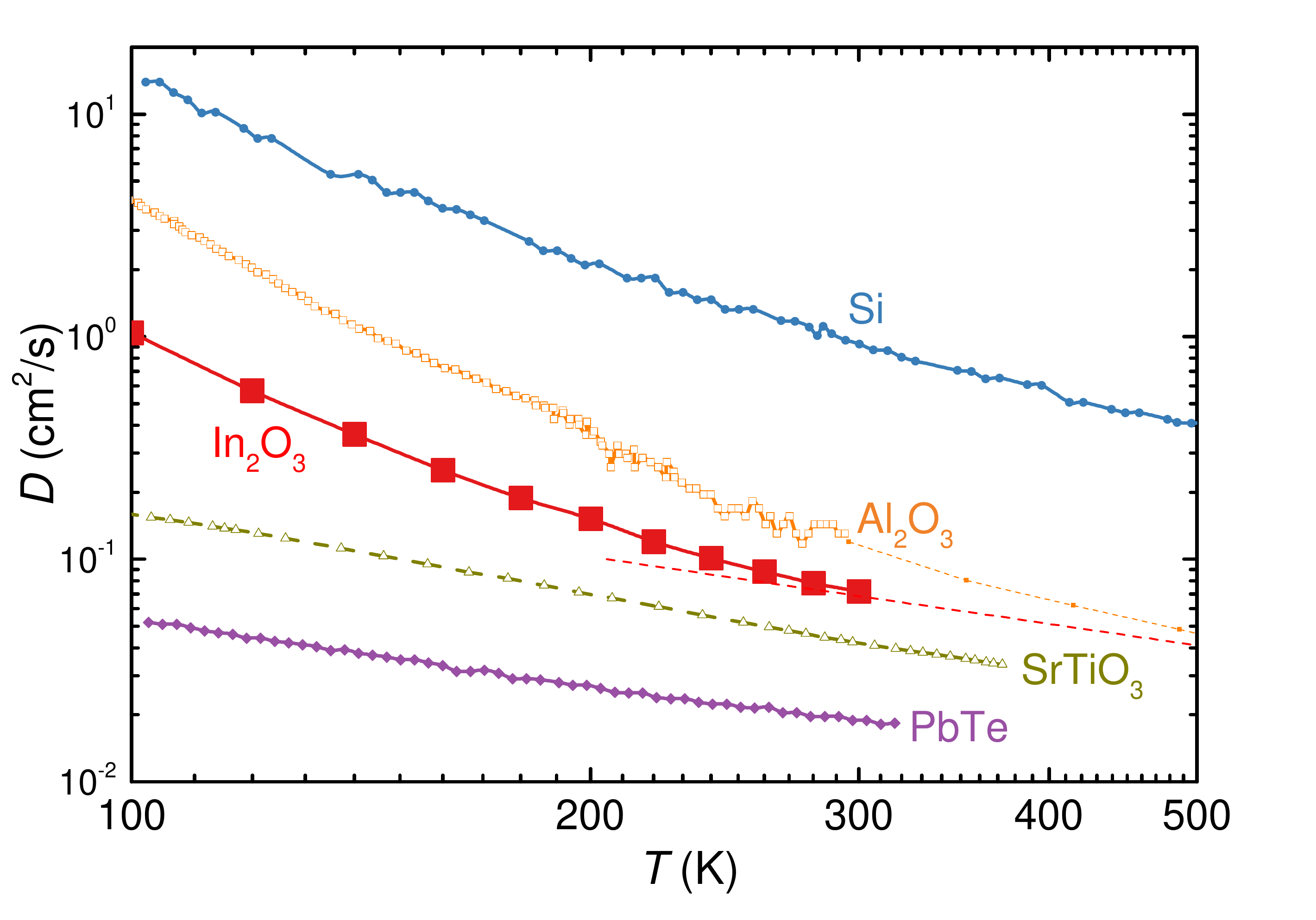}
    \caption{Thermal diffusivity in the intrinsic regime: (a) Temperature dependence of thermal diffusivity, $D$, in In$_2$O$_3$, compared to a few representative solids (Si, SrTiO$_{3}$,  PbTe )~\cite{Martelli2018} and Al$_2$O$_3$\cite{Cahill1988,Hofmeister2014}. The dashed line represents ~$T^{-1}$. At room temperature, $D$ in In$_2$O$_3$ (as well as in  Al$_2$O$_3$ and in Si) approaches this behavior without fully attaining it.}
    \label{fig:Diffusivity}
\end{figure}

Tantalizingly, in sample 6, where the low-temperature  $\ell_{ph}$ is close to the sample thickness and phonons become ballistic,  $\ell_{ph}$(T) peaks at 10 K, a possible signature of phonon hydrodynamics~\cite{beck1974}. In a handful of solids~\cite{beck1974,Martelli2018,machida2018,Machida309}, the temperature dependence of the phonon mean free path displays a local maximum followed by a local minimum. This has been attributed to a regime, first identified by Gurzhi~\cite{Gurzhi1968}, where normal scattering among phonons prevails over both ballistic scattering and umklapp scattering leading to the so-called Poiseuille flow~\cite{beck1974}. It is too early, however, to conclude at this stage that the peak in $\ell_{ph} (T)$ is a Poiseuille peak. No clear Knudsen minimum is visible down to 5 K. Moreover, the thermal conductivity of the best sample may also be affected by extended disorder, since  the phonon wavelength at 10 K is still "only" 30 nm long. This peak is absent in sample 5, where $\ell_{ph}$ falls well below the crystal thickness. Note that above 100 K, $\ell_{ph}$ of all samples merge and approach a $T^{-1}$ behavior. At room temperature, $\ell_{ph}$ becomes as short as 4 nm,  still four times longer than the very long lattice parameter. We come back to this intrinsic regime below.

\begin{figure*}
    \centering
    \includegraphics[width=17cm]{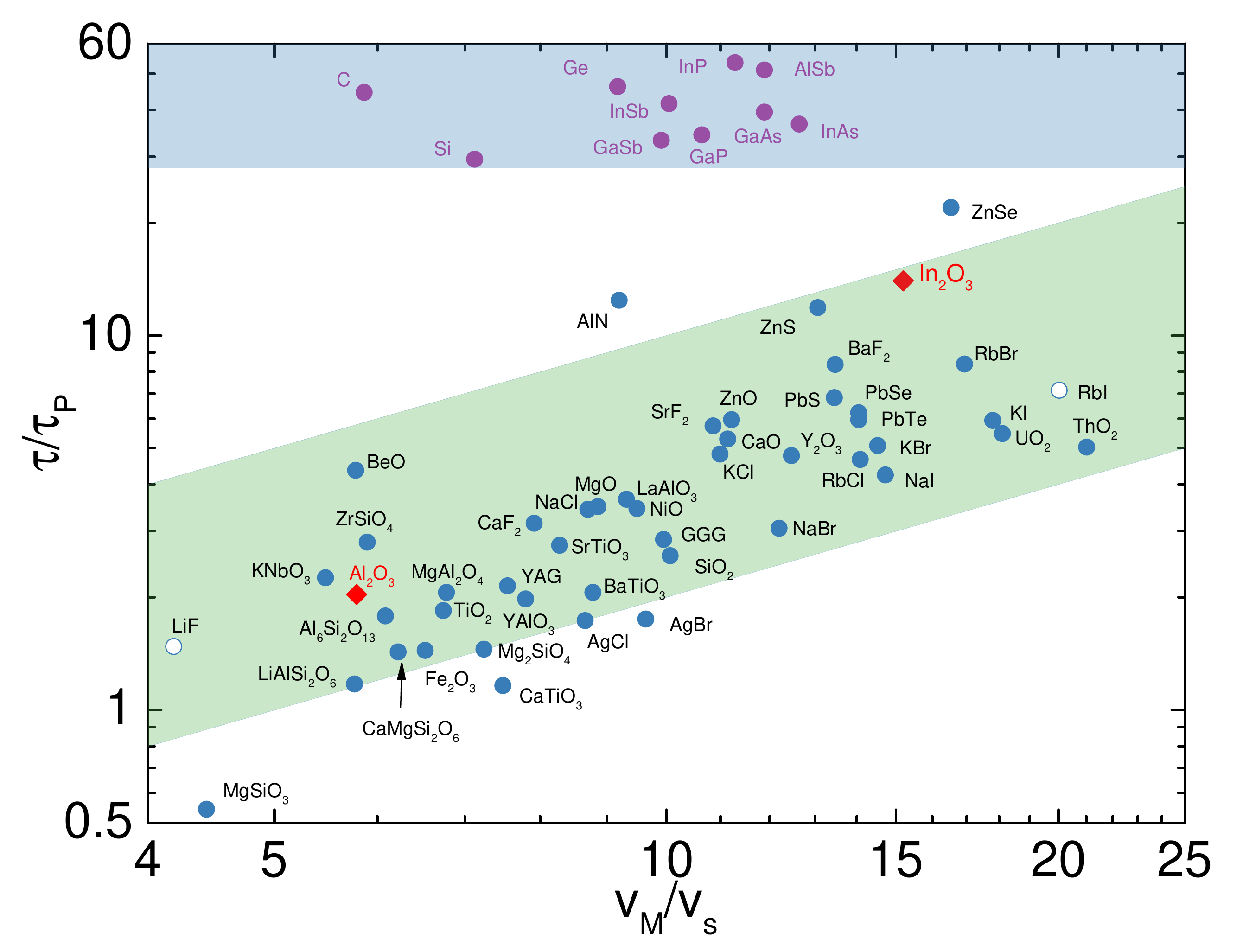}
    \caption{Scattering time, Planckian time, sound velocity, and melting velocity:  The $\tau/\tau_{P}$ ratio vs the $v_{M}/v_{s}$ ratio in different systems adopted from Ref~\cite{Mousatov2020} with the new data point for In$_{2}$O$_{3}$. Note that $\tau$  in In$_{2}$O$_{3}$  was derived at room temperature. Its value at higher temperature may be slightly lower. The blue region contains column IV elements and   III-V semiconductors, which do not participate in this trend. Most other systems lie in a green stripe whose boundaries are set by $0.6\pm 0.4 v_{M}/v_{s}$. Comparisons between In$_{2}$O$_{3}$  and Al$_{2}$O$_{3}$ and between LiF and RbI reveal the role played by atomic mass in setting the sound velocity (see text). }
    \label{fig:tau}
\end{figure*}


Figure \ref{fig:Kappa_campare} compares the thermal conductivity of the best In$_2$O$_3$ sample (sample 6) with a selection of other insulators. In all, $\kappa$ peaks at an intermediate temperature. The peak observed in In$_2$O$_3$ (5000 WK$^{-1}$m$^{-1}$ at 20 K)  is smaller than what is reported in diamond (30000 WK$^{-1}$m$^{-1}$ at 60 K)~\cite{Inyushkin2018} and in sapphire~\cite{Cahill1988} and slightly exceeds the peak in silicon~\cite{Glassbrenner1964}. Remarkably, the crystal structure of In$_{2}$O$_{3}$ is quite distinct from either the diamond structure or the $\alpha$-Al$_{2}$O$_{3}$ structure. It has two nonequivalent sites for indium atoms~\cite{Bierwagen_2015,Karazhanov2007}. An indium atom is surrounded by oxygen atoms either in an octahedral coordination or in a trigonal prismatic coordination. In both cases, indium has six immediate oxygen neighbors. In the octahedral coordination all these O atoms are equidistant from the In atom. In the trigonal prismatic coordination, however, there is a slight difference in distance between three pairs of O atoms. Each oxygen atom is surrounded by one indium atom of the first type and three indium atoms of the second type~\cite{Reunchan2011}.  Thus, the octahedral coordination is present in both sapphire and indium oxide, two binary compounds with remarkably large low-temperature thermal conductivity. We note that other insulators such as Tl$_2$O$_3$~\cite{Glans2005} crystallize in cubic bixbyite structure, but their thermal conductivity remains unexplored. Let us also note that in $\beta$-Ga$_2$O$_3$, $\kappa$ peaks to 560 WK$^{-1}$m$^{-1}$~\cite{Handwerg_2015}. 
This is to be contrasted with ternary oxides, such as perovskites~\cite{Tachibana}, where $\kappa_{peak}$ is in the range of 40 WK$^{-1}$m$^{-1}$, two orders of magnitude smaller than  the two binary oxides of Fig. \ref{fig:Kappa_campare}. As seen in the figure, and discussed in more detail below, this  difference disappears at room temperature. 


Let us now turn our attention to the magnitude of thermal diffusivity in the intrinsic regime. Above 100 K, thermal conductivity in all six samples becomes identical in magnitude and in temperature dependence. Heat conduction in this temperature range is governed by phonon-phonon scattering. Our data combined with specific heat reported previously~\cite{Cordfunke1992} allow us to extract the thermal diffusivity ($D=\frac{\kappa}{c'}$, where $c'$ is the specific heat per volume). As seen in Fig. \ref{fig:Diffusivity}, $D$ evolves faster than $T^{-1}$ below 250 K, but becomes close to the expected $T^{-1}$ behavior near room temperature. The figure compares In$_{2}$O$_{3}$ with a few other solids. The room-temperature thermal diffusivity of In$_{2}$O$_{3}$ is only 7 mm$^2$s$^{-1}$. This is more than one order of magnitude lower than Si ($D\approx 90$ mm$^2$s$^{-1}$), but close to what has been observed in SrTiO$_3$ ($D\approx 4$ mm$^2$s$^{-1}$)~\cite{Martelli2018} and other ternary oxides~\cite{Hofmeister2010,Hofmeister2014} including cuprates~\cite{Zhang2019}.

In this intrinsic regime, the asymptotic $T^{-1}$ temperature dependence of thermal diffusivity represents the temperature dependence of the phonon-phonon scattering cross section. Recently, an empirical lower boundary to  $D$ in this regime was noticed~\cite{Martelli2018,Behnia_2019}. Expressing thermal diffusivity, $D$, as a function of the Planckian time,  $\tau_{P}=\hbar/k_{B}T$, one can write 
\begin{equation}
    D = s v^{2}_{s} \tau_{P}.
\end{equation}

Here, $s$ is a dimensionless parameter (representing the ratio of the  scattering time $\tau$ to $\tau_{P}$). It was observed~\cite{Behnia_2019} that $s$ does not fall below unity in any known solid. According to our data, $s$ $\approx$ 14 in In$_{2}$O$_{3}$ at room temperature. Note that  since $D$ is still evolving faster than $T^{-1}$ below 300 K, our data can only put an upper boundary to $s$. In the case of Al$_{2}$O$_{3}$, data above room temperature~\cite{Hofmeister2014} reduce $s$, since the slope of the $T^{-1}$ thermal diffusivity evolves between 300 and 1200 K. Thus, In$_{2}$O$_{3}$, in spite of being an excellent conductor of heat at low temperatures, emerges more akin to low-diffusivity oxides~\cite{Martelli2018,Zhang2019,Hofmeister2010}.

A recent theoretical account of the empirical boundary to thermal diffusivity~\cite{Martelli2018,Behnia_2019,Zhang2019} was put forward by Mousatov and Hartnoll~\cite{Mousatov2020}, who argued that there is a quantum-mechanical bound to the sound velocity in solids,  because the energy required to hold the crystal together and the highest phonon energy are linked. Note that, the classical distribution of phonons above their Debye temperature notwithstanding, the speed of sound in solids can be linked to $\hbar$, as recently demonstrated by Trachenko \textit{et al.}~\cite{Trachenko2020}. Mousatov and Hartnoll~\cite{Mousatov2020} defined a "melting velocity":  $v_{M} = k_{B}T_{M}a/\hbar$ ($T_{M}$ is the melting temperature, $k_B$ is the Boltzmann constant, and $a$ is the interatomic distance). In their picture, the  smaller $v_{M}/v_{s}$, the closer would be the system to the Planckian bound to dissipation and the lower should be the $\tau/\tau_{P}$ ratio. Their examination of available data on numerous solids (see Fig. \ref{fig:tau}, adapted from Ref. \cite{Mousatov2020}) confirmed this conjecture. As seen in the figure, most solids cluster in the green stripe, where $\tau/\tau_{P}$ is roughly proportional to $v_{M}/v_{s}$.  The list includes various families of ionic salts as well as binary and ternary oxides (in which as we saw above the magnitude of $\kappa_{peak}$ can differ by several orders of magnitude).  On the other hand, as previously noticed~\cite{Mousatov2020}, adamantine solids  (in the blue region) do not follow this trend. One may speculate that the three-phonon scattering phase space~\cite{Lindsay_2008} in these solids is  too small for the relevance of the Planckian bound.

In the Mousatov-Hartnoll plot~\cite{Mousatov2020}, the vertical and horizontal axes are experimentally measurable quantities and there is a visible correlation across  materials. Plots showing correlations between two structural properties across materials  were reported by Grimvall and Sj\"odin~\cite{Grimvall_1974}. These plots are to be considered in the context of the impressive success of the Lindemann criterion for melting in spite of its shortcoming, namely, an exclusive focus on the solid phase and the neglect of the energetics of its liquid competitor~\cite{Wallace1991}.

We can add In$_{2}$O$_{3}$ to this plot. It has a melting temperature of $T_M=2223$ K~\cite{Galazka2013a}, an interatomic distance of $a=0.21$ nm~\cite{Karazhanov2007}, and a sound velocity of $v_{s}$ = 4.4 km/s ~\cite{Ashida2009}. As a result, one finds $v_M/v_s\approx 15$. Our room-temperature data set an upper bound to  $\tau/\tau_{P}\leq$ 15. As seen in Fig. \ref{fig:tau}, these numbers put In$_{2}$O$_{3}$ inside the green stripe in company of numerous other oxides and ionic salts. According to Ref. \cite{Preissler2013}, the sound velocity is 6.4 km/s. Taking this value instead of 4.4 Km/s \cite{Ashida2009} would change both ratios, but it would keep In$_2$O$_3$ inside
the green strip.

Interestingly, In$_{2}$O$_{3}$ and Al$_{2}$O$_{3}$ reside at the opposite ends of the green stripe. This is understandable. Mousatov and Hartnoll~\cite{Mousatov2020} already noticed that what puts two alkali halides (LiF and RbI) with identical crystal structures (and comparable  room-temperature thermal conductivities) far away in this plot is the difference in the atomic masses. Given that the atomic mass of In (115) is about four times larger than Al (27), the same can be said about In$_{2}$O$_{3}$ and Al$_{2}$O$_{3}$. In both cases,  heaviness reduces the sound velocity~\cite{Trachenko2020} and lightness promotes zero-point motion and proximity to the Planckian boundary~\cite{Mousatov2020}. Note, however, that In$_{2}$O$_{3}$ and Al$_{2}$O$_{3}$ have different crystal structures.

In summary, we measured the thermal conductivity of the transparent semiconductor In$_{2}$O$_{3}$ and found a peak at 20 K whose amplitude is surpassed only by a few solids. The peak thermal conductivity was found to be extremely sensitive to disorder. In the intrinsic regime, thermal diffusivity respects a boundary found in other insulators. 

\section{ACKNOWLEDGMENTS}
We thank Sean Hartnoll and Kostya Trachenko for discussions and Detlef Klimm  for a critical reading of the manuscript. This work was supported in France by the Agence Nationale de la Recherche  (Grants No. ANR-18-CE92-0020-01 and No. ANR-19-CE30-0014-04), and by Jeunes Equipes de l$'$Institut de Physique du Coll\`ege de France and by a grant attributed by the Ile de France regional council. It was partly performed in the framework of GraFOx, a Leibniz-Science Campus partially funded by the Leibniz Association – Germany. It was also supported by the National Key Research and Development Program of China (Grant No. 2016YFA0401704), the National Science Foundation of China (Grants No. 51861135104 and No. 11574097) and Fundamental Research Funds for the Central Universities (Grant No. 2019kfyXMBZ071). L. X. acknowledges a Ph.D. scholarship by the China Scholarship Council (CSC).

\bibliography{apssamp}
\end{document}